\documentclass{article}
\usepackage{spconf,amsmath,graphicx, amsfonts, amsthm, xcolor}
\usepackage{booktabs, subcaption, multirow}
\usepackage{enumitem}

\title{Inferring the Graph Structure of Images for Graph Neural Networks}
\name{Mayur S Gowda$^1$, John Shi$^1$, Augusto Santos$^2$, and Jos\'e M.~F.~Moura$^1$ \thanks{This work is partially supported by NSF grant CCN~2327905.\\
mayursri@andrew.cmu.edu, jshi3@andrew.cmu.edu, \\augusto.santos@lx.it.pt, moura@ece.cmu.edu
}
\address{$^1$Department of Electrical and Computer Engineering, Carnegie Mellon University, Pittsburgh, PA\\
$^2$Instituto de Telecomunica\c{c}\~{o}es-IT, Lisbon, Portugal 
}}

\begin{document}
\ninept
\maketitle
\begin{abstract}
Image datasets such as MNIST are a key benchmark for testing Graph Neural Network (GNN) architectures. The images are traditionally represented as a grid graph with each node representing a pixel and edges connecting neighboring pixels (vertically and horizontally). The graph signal is the values (intensities) of each pixel in the image. The graphs are commonly used as input to graph neural networks (e.g., Graph Convolutional Neural Networks (Graph CNNs) \cite{kipf, du2018topology}, Graph Attention Networks (GAT) \cite{GAT}, GatedGCN \cite{gatedgcn}) to classify the images.

In this work, we improve the accuracy of downstream graph neural network tasks by finding alternative graphs to the grid graph and superpixel methods to represent the dataset images, following the approach in \cite{augusto1, augusto2}. We find row correlation, column correlation, and product graphs for each image in MNIST and Fashion-MNIST using correlations between the pixel values building on the method in \cite{augusto1, augusto2}. Experiments show that using these different graph representations and features as input into downstream GNN models improves the accuracy over using the traditional grid graph and superpixel methods in the literature.
\end{abstract}

\textbf{Keywords}: Graph Structure, Images, Graph Neural Networks, Graph Classification, Product Graph

\section{Introduction}
Graph Neural Networks (GNNs) extend traditional deep learning on grid-based signals to signals defined on irregular structures (graphs). Graph Convolutional Neural Networks (Graph CNNs) \cite{kipf, du2018topology} extend the traditional CNN framework, used for image classification, to graph data.

\begin{figure}
    \centering
    \includegraphics[width=0.8\linewidth]{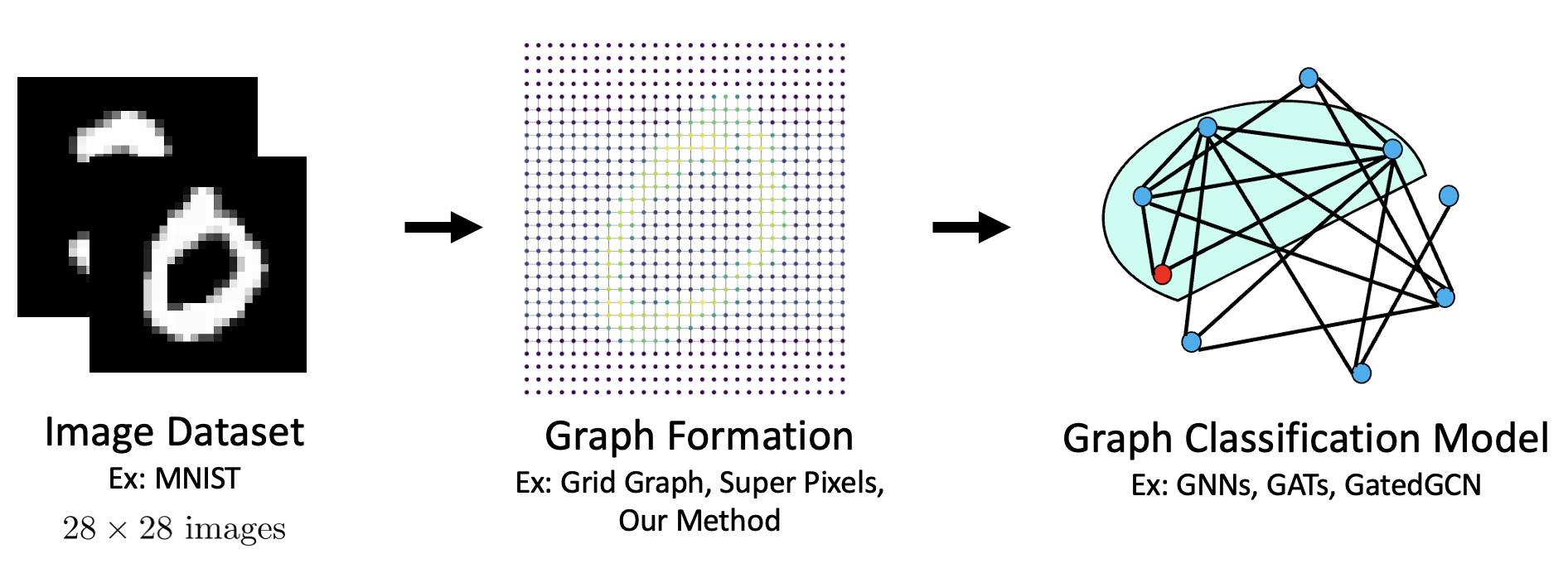}
    \caption{The general Graph Neural Network Pipeline for image classification. Our method engineers a novel graph representation from the image. The graph is then passed to a downstream graph classification model for image classification.}
    \label{fig:summary}
\end{figure}

While designed for graph data, GNNs frequently use image datasets such as MNIST to benchmark GNN architectures. The general GNN pipeline for image classification is shown in Fig. \ref{fig:summary}. In order to classify images using GNNs, we must first represent the image as a graph and the corresponding pixel values and image features as a graph signal. For graph formation, we must consider how to choose and form the 1) nodes, 2) edges to connect the nodes, and 3) features (graph signal). A common, intuitive approach is to represent the image as a grid graph where each node represents a pixel in the image and edges connect adjacent (vertical and horizontal) pixels. Another common approach is using superpixels to form the graph \cite{graphimages}. Depending on how one represents the image using the graph (without changing the GNN model), the performance of GNNs (with fixed architecture) is quite sensitive to the underlying (input) graph representation \cite{graphimages}. After the graph and features are formed, they are passed as input into a downstream Graph Classification Method such as GCN\cite{kipf}, Graph Attention Networks \cite{GAT}, and GatedGCN \cite{gatedgcn} to perform graph classification.

Reference \cite{graphimages} conducted a systematic study of how different graph representations of images using superpixels affect a simple graph CNN model. In terms of the pipeline (shown in Fig. \ref{fig:summary}), they used superpixel methods for graph formation, before classifying using GCN as their downstream graph classification model. 
For graph formation using superpixels, there are many different approaches on how to form nodes, edges, and features \cite{graphimages}:
\begin{enumerate}
    \item \textbf{Nodes:} Each superpixel is a node in the graph. There are many ways to form superpixels from an image with SLIC \cite{slic} being the most common one. The number of superpixels is a hyperparameter, resulting in different densities of pixels per node and different size graphs.
    \item \textbf{Edges:} Once the nodes are formed, we must determine the edges that connect them. Many approaches exist for forming edges such as fully connected graphs, region adjacency graphs (RAGs), and K-Nearest Neighbors with different distance metrics based on spatial and color information \cite{graphimages}.
    \item \textbf{Features:} After the graph is formed, the graph signal (image features) must be chosen. Common features \cite{graphimages} are geometric centroid, number of pixels, the RGB/HSV color average and standard deviation for each superpixel.
\end{enumerate}

Reference \cite{graphimages} evaluated graph representations based on the degree of image segmentation (number of nodes in the graph), the features at each node (the graph signal), and the number of edges between node pairs. Reference \cite{graphimages} concluded that the accuracy of the GCN model \cite{kipf} increases with more detailed representations (with diminishing returns). By increasing the number of nodes (superpixels) in the image and forming graph neighborhoods by connecting only nodes of similar features, the GCN accuracy increases. Reference \cite{graphimages} found that the most significant accuracy increase came from adding the spatial information to the superpixels and a decrease in performance was caused by increasing the size of each node's neighborhood.

In this paper, we focus on the graph formation step in the GNN Pipeline for image classification, forming a graph from the given image dataset (see Fig. \ref{fig:summary}).
We build on the method described in \cite{augusto1, augusto2} to find the graph representation of the image. The method, as presented in \cite{augusto1, augusto2}, infers the underlying graph for networked dynamical systems (NDS) with time series data, where each node is a system whose dynamics are coupled with the dynamics of other neighboring nodes. The method calculates the correlations between (lagged) time series at each node. Then, using these correlations at different lags as the input features, a graph is formed using K-means clustering to find the edges between the nodes. The K-means clustering classifies every possible undirected edge $(i,j)$ between node $i$ and node $j$ as existing in the graph or not existing in the graph. The graph captures the fundamental dependencies among the time series. With the time series considered in \cite{augusto1, augusto2}, the method finds the underlying causal network/graph, which is of ``fundamental importance in downstream tasks" and can be used in many downstream machine learning pipelines.

Similar to networked dynamical systems in \cite{augusto1, augusto2}, pixels in an image may be coupled to other pixels in the image. This is shown in \cite{graphimages}, where GCN performs well for graphs that connect nodes of similar features. We use the method in \cite{augusto1, augusto2} to form graphs for each image. Then, we use these graphs to classify the images using GNNs. 

Our graph formation method has some advantages over superpixels. We do not need to choose the number of nodes. Instead of forming edges based on nearest neighbors, our method connects pixels based on lagged correlation, connecting pixels that have similar features. This allows our method to exclude unimportant background pixels, while connecting important pixels. Superpixels combine neighboring pixels using various methods, e.g., a common superpixel method SLIC \cite{slic} uses a variant of K-means to combine pixels. Unlike superpixels, our method uses all the pixels, and uses K-means to connect the pixels, connecting similar pixels and  disconnecting other pixels from the graph based on feature lagged correlations. Experiments show that our method outperform superpixels in terms of accuracy on downstream graph classification tasks.

For the graph classification model in the general GNN pipeline, we use three different GNNs for graph classification, GCNs, GATs, GatedGCN. GCNs \cite{kipf} operate by aggregating information from neighboring nodes equally, assigning the same importance to all nodes. However, this assumption does not always hold true in real-world graphs. In contrast, instead of using a fixed averaging mechanism (as in GCNs), Graph Attention Networks (GATs) \cite{GAT} introduce an attention mechanism to assign different importance levels to neighboring nodes based on their relevance. This results in more adaptive and meaningful node feature aggregation, leading to improved classification performance. GatedGCN \cite{gatedgcn} improves on GCNs by adding six common techniques: edge feature integration, normalization, dropout, residual connections, feed-forward networks and position encoding, resulting in accuracy improvements over traditional GCNs.

\section{Our Model}
In this section, we outline the parts of our model. We first produce a row and column correlation graph using the method in \cite{augusto1, augusto2}. Then, using these graphs, we form the product graph. 

Then, we focus on the features. We start with the pixel features, then use common image features such as mean and variance, and finally we design our own features. We test these various graphs and features, showing that the Cartesian product graph with correlation features performs the best on image classification for GCNs and GATs in Table \ref{tab:graph_comparison}.

\subsection{Row Graph}
\begin{figure}
    \centering    \includegraphics[width=0.8\linewidth]{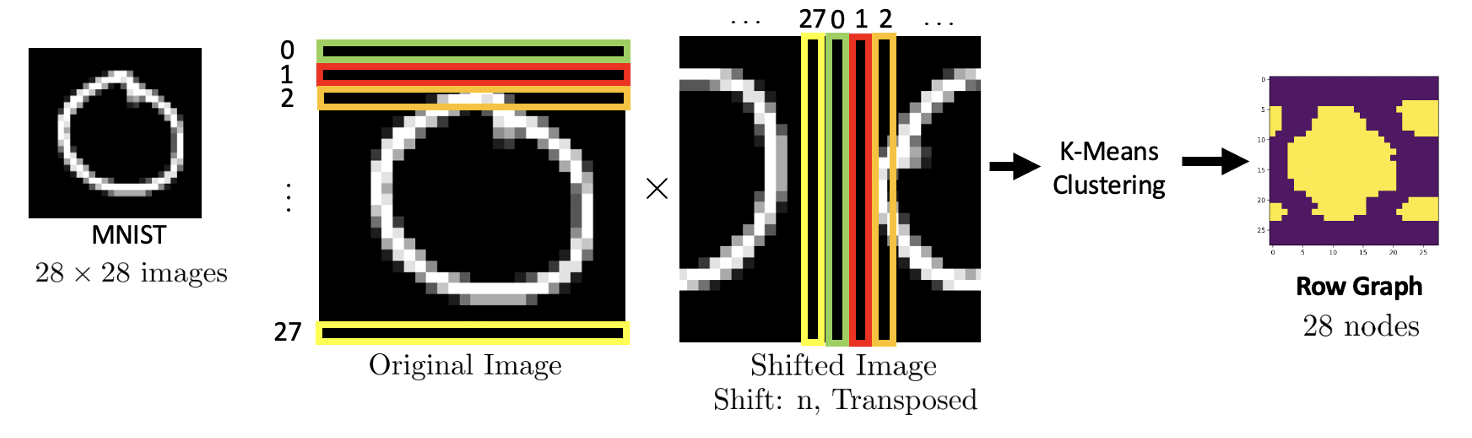}
    \caption{Our approach, building on \cite{augusto1, augusto2}, deployed on MNIST images, to form the row graph. The correlation matrices are found for each lag, then K-means is used to find edges. The lag varies from $n=0$ (no lag) to 27.  }
    \label{fig:rowcorr}
\end{figure}

To produce the row and column graph for each image, we use the method in \cite{augusto1, augusto2}. The process is illustrated in Fig. \ref{fig:rowcorr}.

\begin{figure}
    \centering  \includegraphics[width=0.8\linewidth]{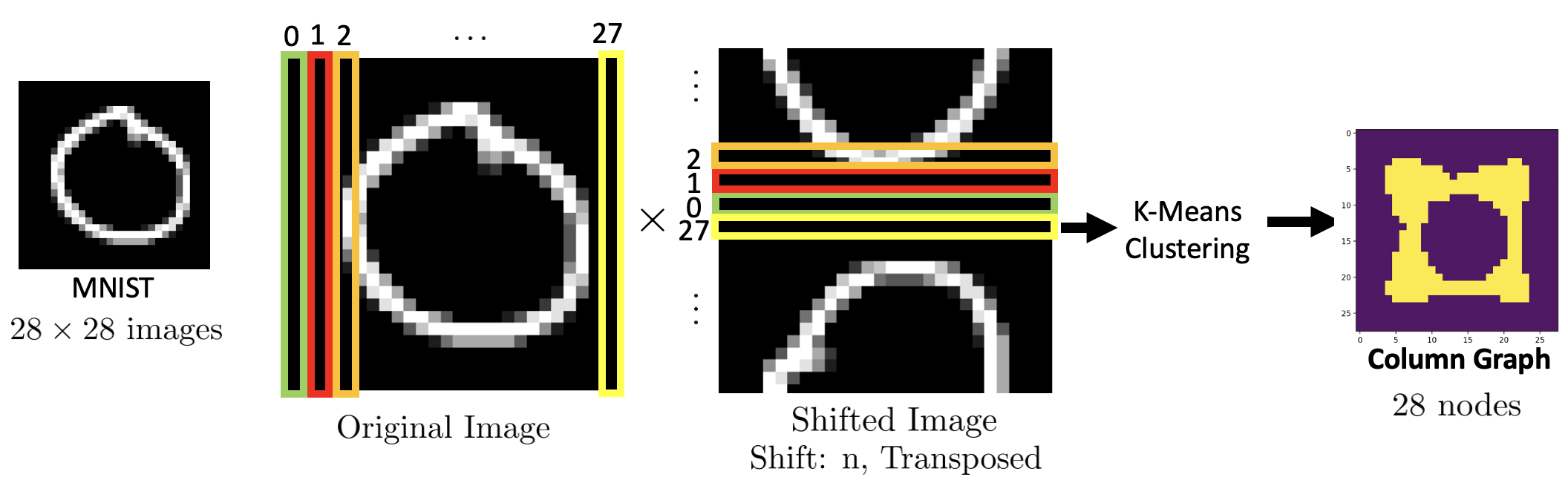}
    \caption{Our approach to form the column graph. The column graph is produced in a similar way to the row graph, but considers the columns of the image instead of the rows. The lag varies from $n=0$ (no lag) to 27. }
    \label{fig:enter-label}
\end{figure}

Let the $N \times N$ image be $A$. For MNIST and Fashion-MNIST, $N = 28$.
Let $A_n'$ be the lagged version of the image $A$ by $n$, $n = 0, 1, \hdots, N-1$. The lagged version, $A_n'$ is formed by circularly shifting the rows of $A$, $n$ times. \begin{equation}
    A_n' = C^n A
\end{equation} where $C$ is cyclic shift matrix.

For each lag, $n = 0, 1, \hdots, N-1$, we calculate the row correlation matrix $r_n$:
\begin{equation}
    r_n = \frac{1}{N} A \cdot A_n'^T
\end{equation} where $T$ is the matrix transpose.

We put the correlation matrices $r_0, r_1, \hdots, r_{N-1}$ as input into K-means clustering to find the row graph. 
The adjacency matrix, created using row-lag correlation, results in a 
$N \times N$ matrix, representing a graph with $N$ nodes, corresponding to rows of the image. Edges represent connections between the rows of the image.
The node features we used correspond to the image pixels, where each node's feature vector is derived from the original pixel intensity. Each node represents a row, so each node has a feature vector of size $N \times 1$ (rows of the image).

\subsection{Column Graph}
The column graph is formed in a similar way to the row graph, using the method in \cite{augusto1, augusto2}, but using the columns instead of the rows.
Let $A_{n_c}'$ be the lagged version of the image $A$ by $n$ ( shifting and lagging the columns), $n = 0, 1, \hdots, N-1$, formed by circularly shifting the \textit{columns} of $A$, $n$ times.
\begin{equation}
A_{n_c}' = A \left(C^n\right)^T
\end{equation}

For each lag, $n = 0, 1, \hdots, N-1$, we calculate the column correlation matrix, $c_n$:

\begin{equation}
    c_n = \frac{1}{N} A \cdot A_{n_c}'^T
\end{equation}

The column correlation matrix, $c_0, c_1, \hdots, c_{N-1}$, are found by circularly shifting the columns of $A$. Then, we input them into K-means clustering to find the column graph.

The adjacency matrix, created using column-lag correlation, results in a 
$N \times N$ matrix, representing a graph with $N$ nodes corresponding to the columns of the image. Edges represent connections between the columns of the image.
The node features we used correspond to the image pixels, where each node's feature vector is derived from the original pixel intensity. Each node represents a column, so each node has a feature vector of size $N \times 1$ (columns of the image).

\subsection{Cartesian Product Graph}
While the row and column graph perform better than the grid graph, 
they only relate the rows and columns of the images. Similar to a grid graph, we want a graph that relates the individual \textit{pixels} of the image.

To form a graph that relates the $N^2$ pixels, we use a product graph, the product of smaller graphs. Some common choices for graph products are the Cartesian and Kronecker products. Graph Signal Processing (GSP) \cite{Sandryhaila:13, overview} uses product graphs to represent time-varying graph data, where the graph is the product graph of a space graph and the cyclic time shift. This is intuitively pleasing because each node in the product graph represents a location in space and a specific time sample \cite{gridgsp}.

Similar to this, by taking the product of the row graph and column graph, each node in the product graph represents a specific row and a specific column, i.e., a specific pixel. So, we form the Cartesian product graph using the following equations:
\begin{align}
A_{2} &= A_r \otimes I + I \otimes A_c \\
    A_{\times} &= A_2  \odot \left(A_r \otimes A_c + A_c \otimes A_r \right)
\end{align} where $A_r$, $A_c$, and $A_x$ are the adjacency matrices of the row, column, and product graph respectively, $I$ is the identity matrix, $\otimes$ is the Kronecker product, and $\odot$ is the pointwise (elementwise) product.

\begin{figure}
    \centering    \includegraphics[width=0.4\linewidth]{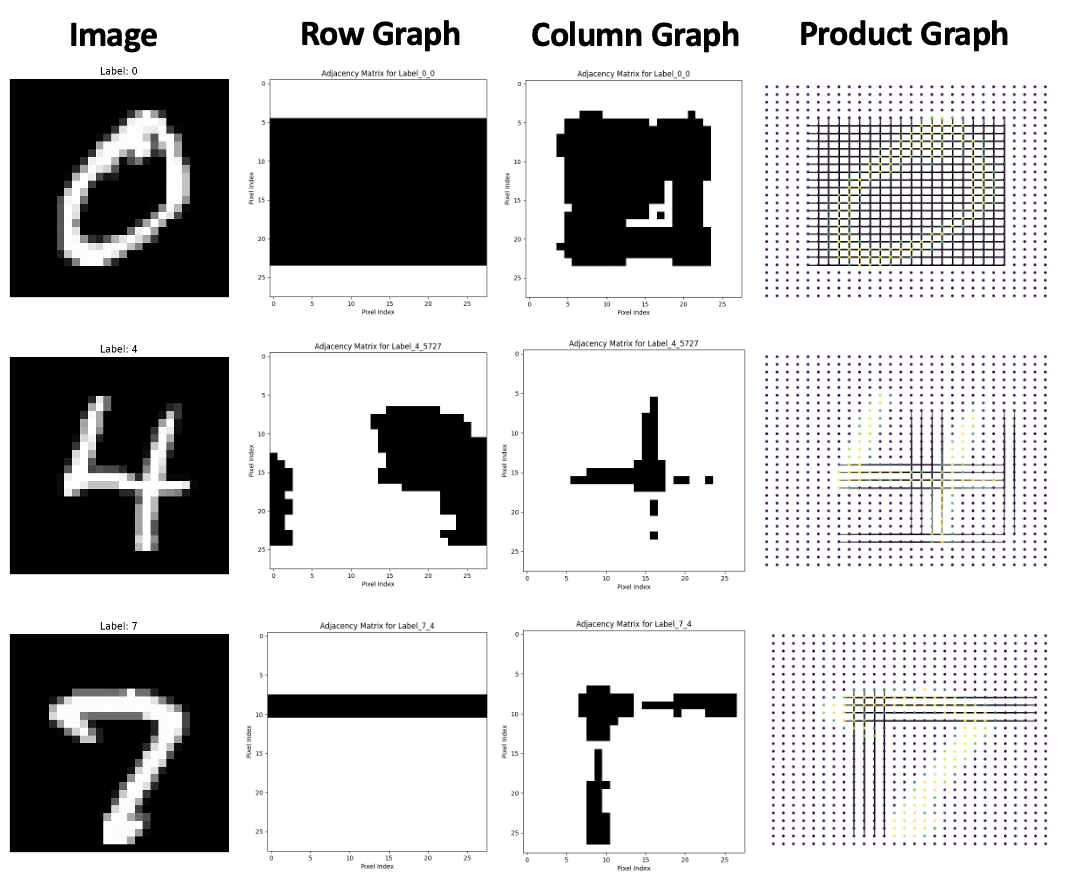}
    \\
    \includegraphics[width=0.4\linewidth]{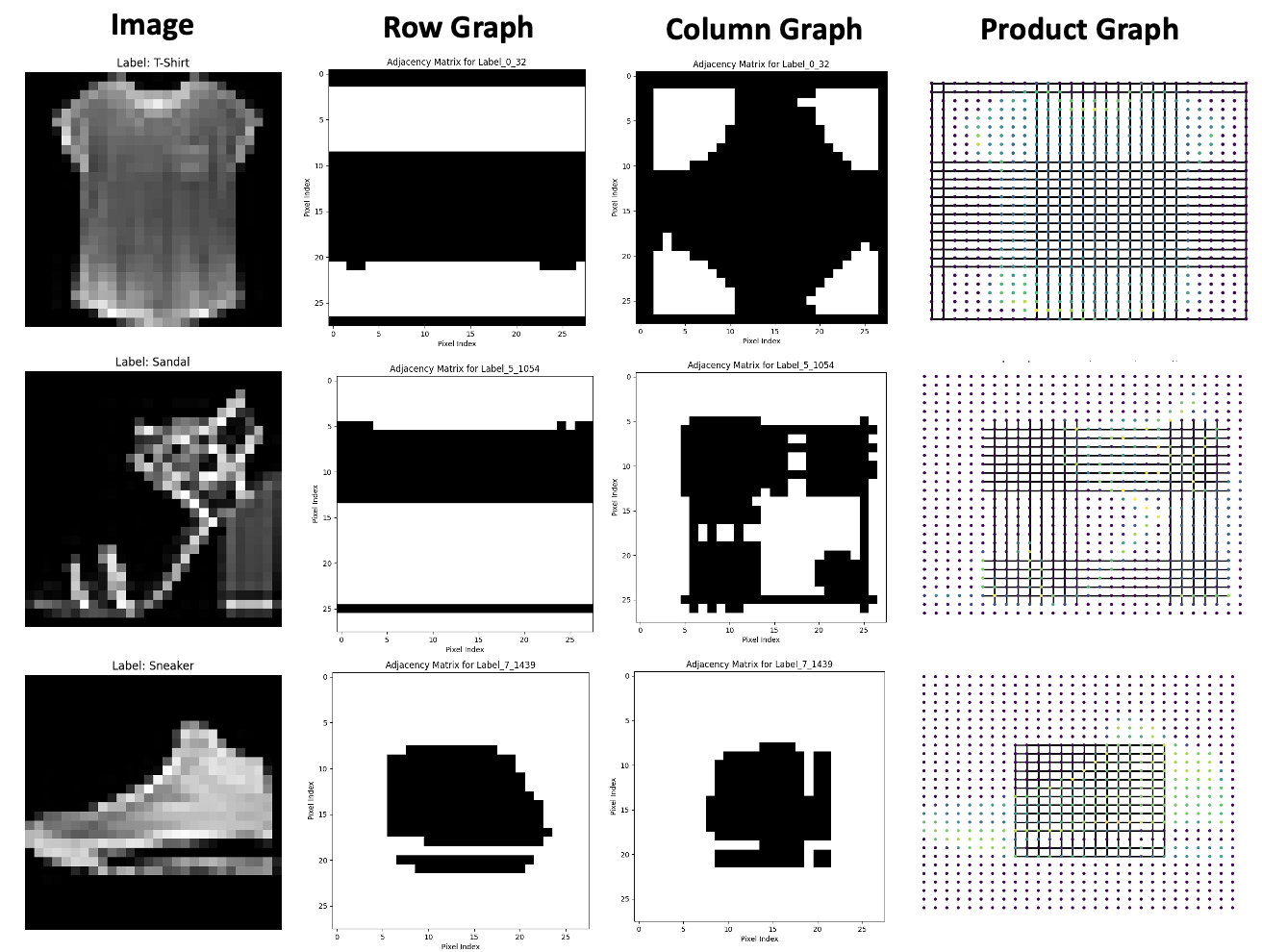}
    
    \caption{Examples of the row, column, product graph formed in Sections 2.1, 2.2, and 2.3 for MNIST (top) and Fashion-MNIST (below).}
    \label{fig:rowcolgraphs}
\end{figure}

Examples of the row, column, and product graph for MNIST and MNIST-Fashion images are shown in Fig. \ref{fig:rowcolgraphs}.

\subsection{Standard Features}

\begin{figure}
    \centering
    \includegraphics[width=0.6\linewidth]{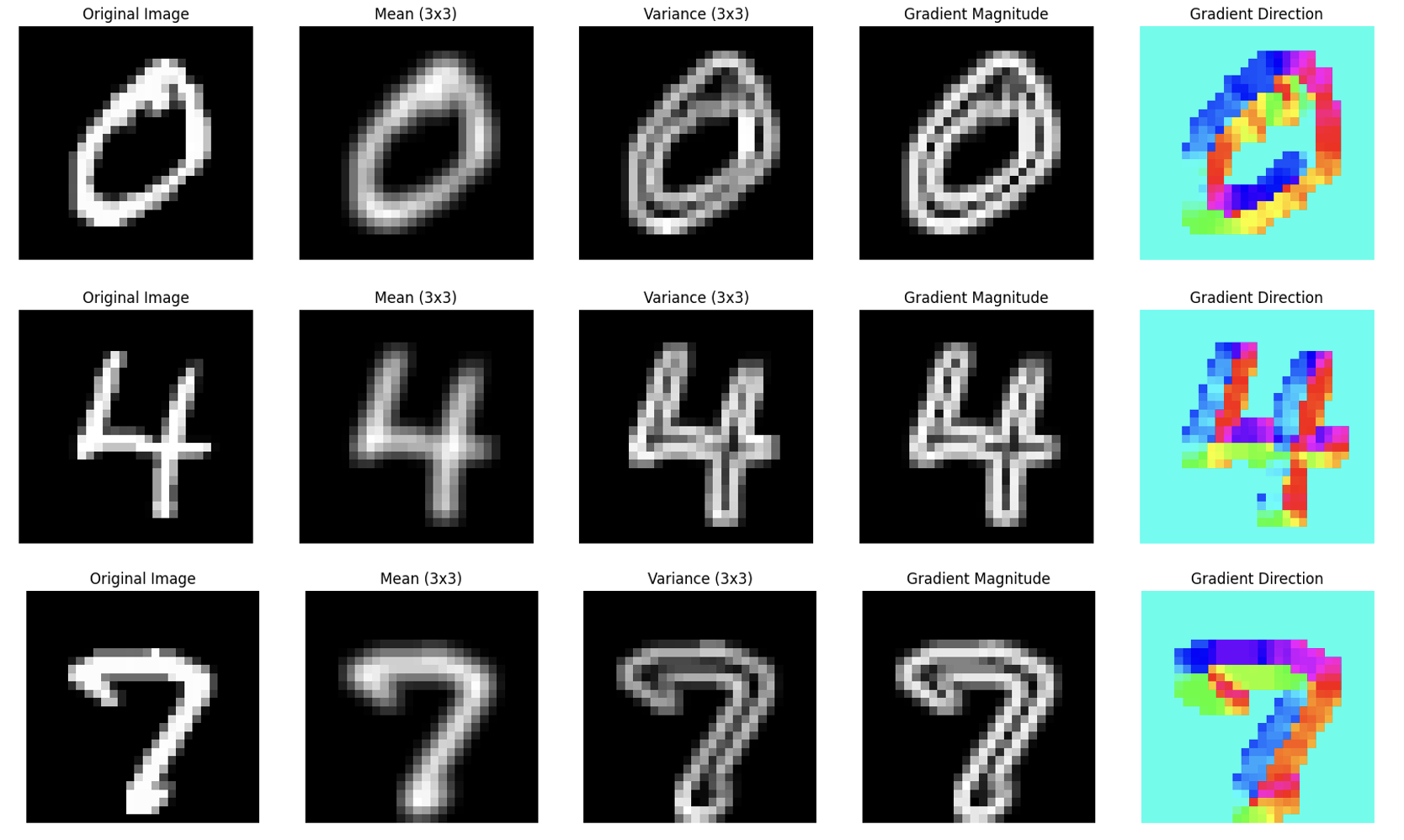}
    \caption{Common image features on MNIST images: Mean, Variance, Gradient Magnitude, and Gradient Direction.}
    \label{fig:standard}
\end{figure}

For the row and column graphs, we used a $N \times 1$ feature vector, which represents the raw pixel intensities along the row/column dimension for each node in the graph. For the product graph of $N^2$ nodes, the feature is a single pixel intensity value.
When using only a single feature per node, the model may have limited information to distinguish nodes beyond their connectivity, making it heavily reliant on the graph structure. In this section and the next section, we consider additional image features. We consider common image features such as mean, variance, gradient magnitude, and gradient direction (of pixels and their neighbors). Examples are in Fig. \ref{fig:standard}.

\subsection{Correlation Features}

We use a cyclic permutation matrix \( C \) to shift the image $A$ and compute lagged versions, $A_l = C^l A$.

For each shift, we compute:

\begin{equation}
F_l = \frac{A + A^T_l}{2}
\end{equation}

This captures how pixel intensities change along structured directions by averaging the original and its transposed shift. To build robustness, we aggregate over all \( N \) lags:

\begin{equation}
\text{NodeFeatureMatrix} = \frac{1}{N} \sum_{l=0}^{N-1} F_l
\end{equation}

We then form a $N^2$-node feature matrix using Kronecker and Cartesian combinations of the row-wise and column-wise NodeFeatureMatrices, $G_r$ and $G_c$. These are the correlation features.

\begin{equation}
G_{\text{mean}} = \frac{G_r \otimes G_c + G_c \otimes G_r}{2} \in \mathbb{R}^{N^2 \times N^2} \end{equation}
where \( col \) and \( row \) are the NodeFeatureMatrices computed from column-wise and row-wise correlations, respectively. This captures both local and long-range structural dependencies between pixels.

Each row of \( G_{\text{mean}} \) serves as a \( (N^2,) \) feature vector, encoding global structural similarity for each pixel. This acts as a graph-based positional embedding derived from correlations.

\begin{figure}
    \centering
    \includegraphics[width=0.2\linewidth]{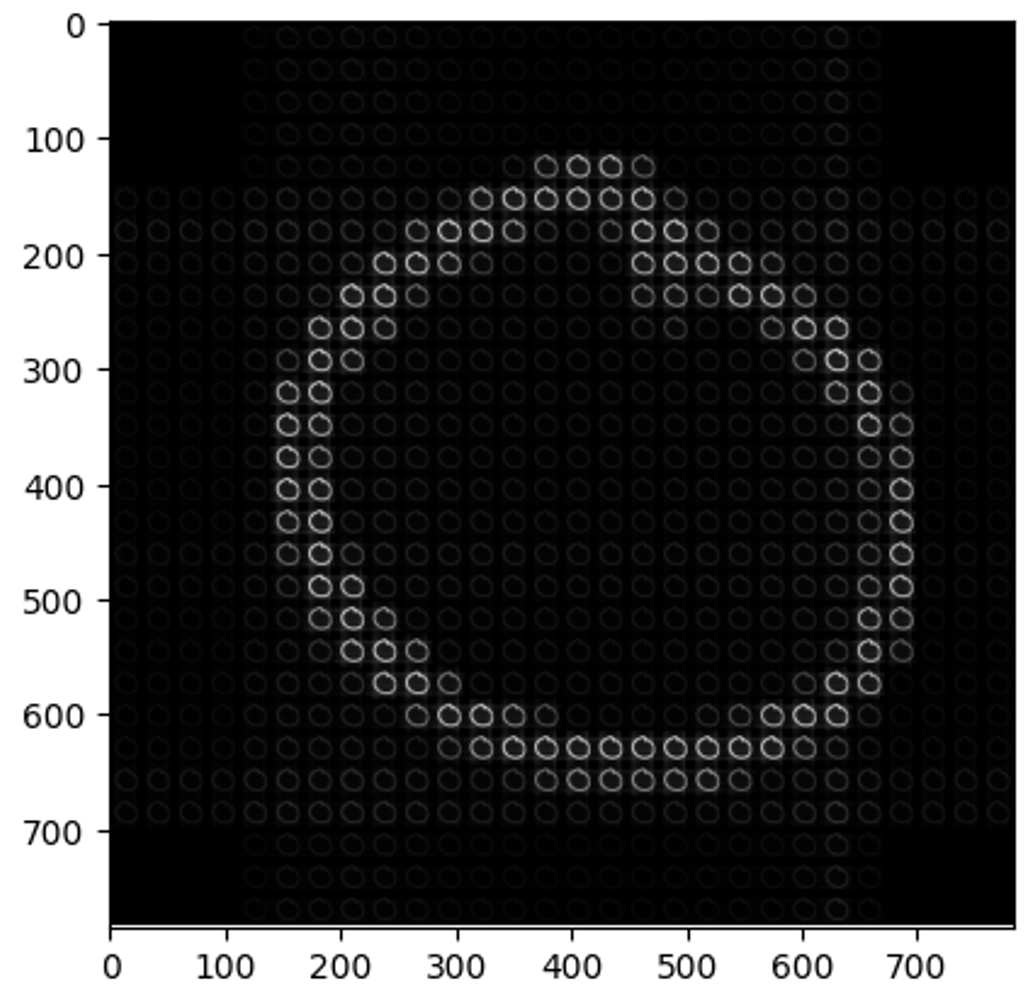}
    \caption{An example of correlation features for MNIST. The features are higher for the parts of the number in the image and are lower for background pixels.}
    \label{fig:corrfeatures}
\end{figure}

An example of correlation features on MNIST is shown in Fig. \ref{fig:corrfeatures}. The correlation features accurately highlight the parts of the number in the image.

\begin{table*}[htb!]
\centering
\renewcommand{\arraystretch}{1} 
\begin{tabular}{|@{\hskip-.5pt}c@{\hskip-.5pt}|c|c|c|c|c|c|c|c|c|c|c|c|}
\hline
\textbf{Dataset} 
& \multicolumn{2}{c|}{\begin{tabular}{c} Grid \\ Graph \end{tabular}} 
& \multicolumn{2}{c|}{\begin{tabular}{c} Row \\ Graph (2.1) \end{tabular}} 
& \multicolumn{2}{c|}{\begin{tabular}{c} Column \\ Graph (2.2) \end{tabular}} 
& \multicolumn{2}{c|}{\begin{tabular}{c} Product Graph \\ (Pixel) (2.3) \end{tabular}} 
& \multicolumn{2}{c|}{\begin{tabular}{c} Product Graph \\ (Common) (2.4) \end{tabular}} 
& \multicolumn{2}{c|}{\begin{tabular}{c} Product Graph \\ (Ours) (2.5) \end{tabular}} \\
\hline
 & \textbf{GCN} & \textbf{GAT} 
 & \textbf{GCN} & \textbf{GAT} 
 & \textbf{GCN} & \textbf{GAT} 
 & \textbf{GCN} & \textbf{GAT} 
 & \textbf{GCN} & \textbf{GAT} 
 & \textbf{GCN} & \textbf{GAT} \\
\hline
MNIST 
& 27.70 & 73.99
& 71.35 & 92.58 
& 87.18 & 95.07 
& \begin{tabular}{c} 55.05 \\ $\pm$ 2.89 \end{tabular} 
& \begin{tabular}{c} 76.45 \\ $\pm$ 0.067 \end{tabular} 
& \begin{tabular}{c} 73.36 \\ $\pm$ 2.04 \end{tabular} 
& \begin{tabular}{c} 78.18 \\ $\pm$ 0.069 \end{tabular} 
& \begin{tabular}{c} 94.97 \\ $\pm$ 0.246 \end{tabular} 
& \begin{tabular}{c} 96.91 \\ $\pm$ 0.016 \end{tabular} \\
\hline
\begin{tabular}{c} Fashion- \\ MNIST \end{tabular}
& 40.29 & 77.30 
& 74.06 & 85.45 
& 80.72 & 86.96 
& \begin{tabular}{c} 34.47 \\ $\pm$ 4.81 \end{tabular} 
& \begin{tabular}{c} 75.43 \\ $\pm$ 0.061 \end{tabular} 
& \begin{tabular}{c} 72.94 \\ $\pm$ 4.058 \end{tabular} 
& \begin{tabular}{c} 79.48 \\ $\pm$ 0.176 \end{tabular} 
& \begin{tabular}{c} 85.10 \\ $\pm$ 0.587 \end{tabular} 
& \begin{tabular}{c} 89.1 \\ $\pm$ 0.067 \end{tabular} \\
\hline
\end{tabular}
\caption{Comparison of GCN and GAT performance across different graph construction methods for MNIST and Fashion-MNIST datasets. Our method's improvements gradually increase the accuracy compared to a grid graph. We found that using both the product graph and correlation features does the best among all the improvements.}
\label{tab:graph_comparison}
\end{table*}
\section{Datasets and Models}
The experiments were conducted using the MNIST and Fashion-MNIST datasets. Both datasets have 70,000 images with 60,000 images for training and 10,000 for testing.
They both consist of $28 \times 28$ size grayscale images. MNIST consists of images of handwritten digits. Fashion-MNIST consists of images of clothing. Three models were used: Graph Convolutional Networks (GCN) \cite{kipf}, Graph Attention Networks (GAT) \cite{GAT} and GatedGCNs  \cite{gatedgcn}. 

The GCN model employs a deep, pyramid-style architecture consisting of seven GCNConv layers
with progressively increasing hidden dimensions (64 → 128 → 256 → 512 → 1024). Each layer is followed by batch normalization and ReLU activation.
For GAT, the model architecture consists of four stacked GATConv layers with 8 attention heads each,
using ELU activations and residual connections. Each layer is followed by layer normalization.
Lastly, GatedGCN was used out of the box, with models directly from https://github.com/LUOyk1999/tunedGNN-G. 

For each model (GCN, GAT, GatedGCN), we performed image classification on MNIST and Fashion-MNIST using 3 different graph formation methods: the grid graph, superpixels (using SLIC) \cite{slic}, and our method: Product Graph with Corrleation Features. 
The results are shown in Table \ref{tab:results}.

\section{Results}

Table \ref{tab:graph_comparison} illustrates the accuracies of GCN and GAT for the different graph formation methods we developed in Section 2 for MNIST and Fashion-MNIST using GCNs and GAT. 
Starting from a grid graph, we iteratively improve the graph structure and the graph features used in the graph formation step (Sections 2.1 to 2.5), obtaining a gradual accuracy increase. We use the best model: product graph with correlation features (2.5) for comparison with grid graph and superpixels.
The column graph performs better than the row graph, showing that the columns contain more information than the rows, while the product graph performs better than both. Using common image features (2.4) over just pixel intensities improves the accuracy. Using our correlation features (2.5) further improves the accuracies.

From Table \ref{tab:results}, we observe an accuracy improvement using our method: product graph using correlation features, over the grid graph and superpixel graph formation methods.
This improvement arises because our method forms graphs that mainly connects pixels that form the object, rather than background pixels (see Fig. 2). In contrast, a grid graph considers all pixels equally, i.e., there is no difference between background pixels and the pixels that form the object. Similarly, superpixels still uses the background pixels, grouping them by proximity.
Our method and superpixels form graphs that consider pixels unequally, while a grid graph considers all pixels equally. Similarly, since GCNs aggregate information from neighboring nodes equally and GATs do not, we see a larger difference in accuracy with GCNs over GATs when using the grid graph vs. the graph from our method. One reason that GatedGCN does better on superpixels for MNIST vs. our model is that the out of the box model used may have been specifically tuned to do well on this problem.

\begin{table}[h!]
\centering
\begin{tabular}{|c|c|c|c|}
\hline
\multicolumn{4}{|c|}{\textbf{Dataset: MNIST}} \\
\hline
\textbf{Graph Type} & \textbf{GCN} & \textbf{GAT} & \textbf{GatedGCN} \\
\hline
Grid Graph & $27.70 \pm 0.86$ & $73.99 \pm 0.95$ & $66.39 \pm 5.53$ \\
Superpixels & $27.72 \pm 0.22$ & $79.78 \pm 0.34$ & $\bf{98.71 \pm 0.13}$ \\
Product Graph (Ours) & $\bf{94.97 \pm 0.24}$ & $\bf{96.91 \pm 0.01}$ & $94.36 \pm 0.02$ \\
\hline
\multicolumn{4}{|c|}{\textbf{Dataset: Fashion-MNIST}} \\
\hline
\textbf{Graph Type} & \textbf{GCN} & \textbf{GAT} & \textbf{GatedGCN} \\
\hline
Grid Graph & $40.29 \pm 5.76$ & $77.30 \pm 0.52$ & $73.71 \pm 0.94$ \\
Superpixels & $30.78 \pm 0.33$ & $80.57 \pm 0.23$ & $75.25 \pm 0.30$ \\
Product Graph (Ours) & $\bf{85.10 \pm 0.58}$ & $\bf{89.10 \pm 0.06}$ & $\bf{86.38 \pm 0.23}$ \\
\hline
\end{tabular}

\caption{Comparison of Various Graph Formation methods on downstream graph classification tasks. Our method does better than the grid graph and superpixels on GCN, GAT, GatedGCN for the MNIST and Fashion-MNIST datasets.}
\label{tab:results}
\end{table}

\section{Conclusion}
In this work, we improve the accuracy for downstream graph classification tasks by choosing a more meaningful graph structure and features. We achieve this by inferring the underlying graph for images using the correlation method in \cite{augusto1, augusto2}. 
Our method uses a product graph, formed by taking the product of row and column graphs. Instead of just using the pixel intensities, we design correlation features. The combination of the product graph and correlation features achieves higher accuracy than the grid, row, column graphs, and with common image features. Our method outperforms the grid graph and superpixel methods for GCN, GAT, and GatedGCN on MNIST and Fashion-MNIST.
Experiments show the accuracy increasing as the graph structure and features become more meaningful.

\newpage
\bibliographystyle{IEEEbib}
\bibliography{strings,refs, refsjournal}

\begin{thebibliography}{10}

\bibitem{kipf}
Thomas~N. Kipf and Max Welling,
\newblock ``Semi-supervised classification with graph convolutional networks,''
\newblock in {\em International Conference on Learning Representations (ICLR)}, 2017.

\bibitem{du2018topology}
Jian Du, Shanghang Zhang, Guanhang Wu, Jose M.~F. Moura, and Soummya Kar,
\newblock ``Topology adaptive graph convolutional networks,''
\newblock {\em CoRR}, 2017.

\bibitem{GAT}
Petar Velickovic, Guillem Cucurull, Arantxa Casanova, Adriana Romero, Pietro Li{\`{o}}, and Yoshua Bengio,
\newblock ``Graph attention networks,''
\newblock in {\em International Conference on Learning Representations {(ICLR)}}, 2018.

\bibitem{gatedgcn}
Yuankai Luo, Lei Shi, and Xiao-Ming Wu,
\newblock ``Can classic gnns be strong baselines for graph-level tasks? simple architectures meet excellence,'' 2025.

\bibitem{augusto1}
Augusto Santos, Diogo Rente, Rui Seabra, and Jos\'{e} M.~F. Moura,
\newblock ``Learning the causal structure of networked dynamical systems under latent nodes and structured noise,''
\newblock in {\em Proceedings of the Thirty-Eighth AAAI Conference on Artificial Intelligence and Thirty-Sixth Conference on Innovative Applications of Artificial Intelligence and Fourteenth Symposium on Educational Advances in Artificial Intelligence}. 2024, AAAI'24/IAAI'24/EAAI'24, AAAI Press.

\bibitem{augusto2}
Augusto Santos, Diogo Rente, Rui Seabra, and José M.~F. Moura,
\newblock ``Inferring the graph of networked dynamical systems under partial observability and spatially colored noise,''
\newblock in {\em ICASSP 2024 - 2024 IEEE International Conference on Acoustics, Speech and Signal Processing (ICASSP)}, 2024, pp. 13156--13160.

\bibitem{graphimages}
Júlia Rodrigues and Joel Carbonera,
\newblock ``Graph convolutional networks for image classification: Comparing approaches for building graphs from images,''
\newblock 05 2024.

\bibitem{slic}
Radhakrishna Achanta, Appu Shaji, Kevin Smith, Aurélien Lucchi, Pascal Fua, and Sabine Süsstrunk,
\newblock ``Slic superpixels,''
\newblock {\em Technical report, EPFL}, 06 2010.

\bibitem{Sandryhaila:13}
A.~Sandryhaila and J.~M.~F. Moura,
\newblock ``Discrete signal processing on graphs,''
\newblock {\em IEEE Trans. Signal Proc.}, vol. 61, no. 7, pp. 1644--1656, April 2013.

\bibitem{overview}
Antonio Ortega, Pascal Frossard, Jelena Kovačević, José M.~F. Moura, and Pierre Vandergheynst,
\newblock ``Graph signal processing: Overview, challenges, and applications,''
\newblock {\em Proceedings of the IEEE}, vol. 106, no. 5, pp. 808--828, 2018.

\bibitem{gridgsp}
John Shi and José M.~F. Moura,
\newblock ``Graph signal processing: The 2d companion model,''
\newblock in {\em ICASSP 2024 - 2024 IEEE International Conference on Acoustics, Speech and Signal Processing (ICASSP)}, 2024, pp. 9806--9810.

\end{thebibliography}

\end{document}